\documentstyle[12pt]{article}
\begin{document}
\def\be{\begin{eqnarray}}
\def\en{\end{eqnarray}}
\def\non{\nonumber}
\def\la{\langle}
\def\ra{\rangle}
\def\ep{\varepsilon}
\def\ums{{\mu}_{_{\overline{\rm MS}}}}
\def\u{\mu_{\rm fact}}
\def\gg{\Delta\sigma^{\gamma G}}
\def\lsim{ {\ \lower-1.2pt\vbox{\hbox{\rlap{$<$}\lower5pt\vbox{\hbox{$\sim$}
}}}\ } }
\def\gsim{ {\ \lower-1.2pt\vbox{\hbox{\rlap{$>$}\lower5pt\vbox{\hbox{$\sim$}
}}}\ } }
\def\dk{\partial\!\cdot\!K}
\def\pr{{\sl Phys. Rev.}~}
\def\prl{{\sl Phys. Rev. Lett.}~}
\def\pl{{\sl Phys. Lett.}~}
\def\np{{\sl Nucl. Phys.}~}
\def\zp{{\sl Z. Phys.}~}

\font\el=cmbx10 scaled \magstep2
{\obeylines
\hfill IP-ASTP-25-95
\hfill December, 1995}

\vskip 1.5 cm

\centerline{\large\bf Sea Quark or Anomalous Gluon Interpretation for
$g_1^p(x)$ ? }
\medskip
\bigskip
\medskip
\centerline{\bf Hai-Yang Cheng}
\medskip
\centerline{ Institute of Physics, Academia Sinica}
\centerline{ Taipei, Taiwan 115, Republic of China}
\bigskip
\bigskip
\bigskip
\bigskip
\centerline{\bf Abstract}
\bigskip
{\small
Contrary to what has been often claimed in the literature, we clarify that
the hard photon-gluon cross section $\gg_{\rm hard}(x)$ in polarized deep
inelastic scattering calculated in the
gauge-invariant factorization scheme does {\it not} involve any soft
contributions and hence it is genuinely {\it hard}. We show that
the polarized proton structure function $g_1^p(x)$ up to the next-to-leading
order of $\alpha_s$ is independent of the
factorization convention, e.g., the gauge-invariant or chiral-invariant
scheme, chosen in defining $\gg_{\rm hard}(x)$ and the quark spin density.
Thereby, it is not pertinent to keep disputing which factorization
prescription is correct or superior.
The hard-gluonic contribution to $\Gamma_1^p$, the first moment of
$g_1^p(x)$, is purely factorization dependent.  Nevertheless, we stress that
even though hard gluons do not contribute to $\Gamma_1^p$ in the
gauge-invariant scheme, the gluon spin component in a proton, which is
factorization independent,
should be large enough to perturbatively generate a negative sea polarization
via the axial anomaly. We briefly comment on how to
study the $Q^2$ evolution of parton spin distributions
to the next-to-leading order of QCD in the chiral-invariant factorization
scheme.

}

\pagebreak

   {\bf 1}.~~Recently we have analyzed the EMC [1] and SMC [2] data
of the polarized proton structure function $g_1^p(x)$ to extract the
polarized parton distributions to the next-to-leading order (NLO) of QCD
by assuming that $Q^2=\la Q^2\ra=10\,{\rm GeV}^2$ for each $x$ bin of the
$g_1^p(x)$ data [3]. Our analysis is performed in two extreme factorization
schemes: gauge-invariant and chiral-invariant ones. In the former
factorization, hard gluons do not contribute to $\Gamma_1^p$, the first moment
of $g_1^p(x)$, and the quark net helicity is $Q^2$ dependent but has a local
gauge-invariant operator expression, whereas gluons do make contributions to
$\Gamma_1^p$ and the chiral-invariant quark spin does not evolve in the
latter scheme. However, we have stressed that physics is independent of the
choice of the factorization prescription; the size of the hard gluonic
contribution to $\Gamma_1^p$ is purely a matter of convention, as first
realized and strongly advocated by Bodwin and Qiu [4] sometime ago.

    Because of the availability of the two-loop polarized splitting
functions $\Delta P^{(1)}_{ij}(x)$ very recently [5], it becomes possible to
embark on a full NLO analysis of the experimental data of polarized
structure functions by taking into account the measured $x$ dependence of
$Q^2$ at each $x$ bin. Two of such analyses are now available. A NLO QCD
analysis is carried out in [6] in the
conventional $\overline{\rm MS}$ scheme, a gauge-invariant
factorization, within the framework of the radiative parton model.
While a sizeable and negative sea polarization is required to describe
all presently available data, the gluon spin density is found to be rather
weakly constrained by the data. On the contrary, a NLO fit presented in [7]
in the chiral-invariant scheme shows that the gluon contribution is large
and positive. Furthermore, the authors of [7] have criticized the
gauge-invariant scheme that it is pathological and inappropriate
because soft contributions are partly included in the hard coefficient
function rather than being factorized into parton spin densities. More
precisely, the hard gluon-photon cross section has the expression [4]
\be
\gg_{\rm hard}(x)_{\rm GI}=\,{\alpha_s\over 2\pi}\left[(2x-1)\left(\ln{Q^2
\over \u^2}+\ln{1-x\over x}-1\right)+2(1-x)\right],
\en
in the gauge-invariant factorization scheme with $\u$ a factorization scale.
An objection to this scheme has been
that the last term in Eq.(1) proportional to $2(1-x)$ appears to arise
from the soft region $k_\perp^2\sim m^2<<\Lambda^2_{\rm QCD}$, and hence
it should be absorbed into the polarized quark distribution [7-9].

   Although the issue of whether or not gluons contribute to $\Gamma_1^p$ was
resolved five years ago that it depends on the factorization convention chosen
in defining the quark spin density  and the hard cross section
for the photon-gluon scattering [4,10],
\footnote{See also a very nice roundtable summary at the {\it Polarized
Collider Workshop} (University Park, PA, 1990) on the theoretical
interpretation of the measured $g_1^p(x)$ [11].}
the fact that the interpretation of $\Gamma_1^p$ is still under
dispute even today and that many recent articles and reviews are still biased
towards or against one of the two popular implications of the measured
$g_1^p(x)$, namely sea quark or anomalous gluon interpretation, demands
a further clarification on this issue. In the present paper, we will point out
that, irrespective of the soft cutoff,
the $2(1-x)$ term in Eq.(1) actually arises from the axial anomaly, i.e.,
from the region where $k_\perp^2\sim\u^2$. Consequently, many criticisms to
the gauge-invariant factorization scheme are in vain.
We will show that even though hard gluons do not contribute to $\Gamma_1^p$ in
the gauge-invariant scheme, the bulk of negative sea polarization must be
perturbatively generated by gluons via the axial anomaly.
We would like to emphasize that none of the results
presented in this paper (it reads like a status review)
are new and they are scattered in the literature.
Nevertheless, we believe that the present paper is useful for clarifing
several confusing issues and for making a unified and consistent
understanding of $g_1^p(x)$.

\vskip 0.3cm
{\bf 2.}~~As far as the first moment of $g_1^p(x)$ is concerned, the
parton-model and OPE approaches are equivalent. However, in order to consider
QCD corrections to $g_1^p(x)$ itself, we shall consider the parton model.
To NLO, the polarized proton structure function is given by
\be
g_1^p(x,Q^2) &=& {1\over 2}\sum^{n_f}_i e^2_i\Big\{\ [\Delta q_i(x,Q^2)+{
\alpha_s(Q^2)\over 2\pi}\Delta f_q(x)\otimes\Delta q_i(x,Q^2)] \non \\
&& +\gg_{\rm hard}(x,Q^2)
\otimes\Delta G(x,Q^2)\Big\},
\en
where $n_f$ is the number of active quark flavors, $\Delta q(x)=q^\uparrow(x)+
\bar{q}^\uparrow(x)-q^\downarrow(x)-\bar{q}^\downarrow(x)$, $\Delta G(x)=
G^\uparrow(x)-G^\downarrow(x)$, and $\otimes$ denotes the convolution
\be
f(x)\otimes g(x)=\int^1_x{dy\over y}f\left({x\over y}\right)g(y).
\en
The $\Delta f_q(x)$ term in Eq.(2) depends on the regularization scheme
chosen, but its first moment is scheme independent at least to NLO:
$\int^1_0\Delta f_q(x)dx=
-2$. Since the polarized photon-gluon cross section $\gg(x)$ has infrared
and collinear singularities at $m^2=p^2=0$ and $k_\perp^2=0$, where $m$ is the
quark mass, $p^2$ is the gluon momentum squared and $k_\perp$ is the quark's
transverse momentum perpendicular to the virtual photon direction, it is
necessary to introduce a soft cutoff. Depending on the infrared regulators,
one obtains
\be
\gg_{\rm CCM}(x,Q^2) &=& {\alpha_s\over 2\pi}\left[(2x-1)\left(\ln{Q^2\over
-p^2x(1-x)}+\ln{1-x\over x}-1\right)+1-2x\right],   \\
\gg_{\rm AR}(x,Q^2) &=& {\alpha_s\over 2\pi}\left[(2x-1)\left(\ln{Q^2\over
m^2}+\ln{1-x\over x}-1\right)+2(1-x)\right],   \\
\gg_{\rm R}(x,Q^2) &=& {\alpha_s\over 2\pi}\left[(2x-1)\left(\ln{Q^2\over
\ums^2}+\ln{1-x\over x}-1\right)+2(1-x)\right],
\en
for the momentum regulator ($p^2\neq 0$) [12], the mass regulator ($m^2\neq
0$) [13], and the modified dimensional regulator ($\ums^2\neq 0$) [14],
respectively. Note that the term $(2x-1)$ in Eqs.(4-6) is nothing but
the spin splitting
function $2\Delta P_{qG}(x)$ and that the term proportional to $2(1-x)$ in (5)
and (6) is an effect of chiral symmetry breaking: It arises from the
region where $k_\perp^2\sim m^2$ in the mass-regulator scheme, and from
$k_\perp^2\sim\ums^2$ in the $n\neq 4~(n>4)$ dimensions
in the dimensional regularization scheme.

The cross sections given in (4-6) are, however, not the desirable perturbative
QCD results since they are sensitive to the choice of the regulator. Although
the $\ln(Q^2/-p^2)$ and $\ln(Q^2/m^2)$ terms, which depend logarithmically
on the soft cutoff, make no contributions to the first moment of $g_1^p(x)$,
it is important to have a reliable perturbative QCD calculation for $\gg(x)$
since we are intertested in QCD corrections to $g_1^p(x)$. To do this, we need
to introduce a factorization scale $\u$, so that
\be
\gg(x,Q^2)=\,\gg_{\rm hard}(x,Q^2,\u^2)+\gg_{\rm soft}(x,\u^2)
\en
and the polarized photon-proton cross section is decomposed into
\be
\Delta\sigma^{\gamma p}(x,Q^2)=\sum_i^{n_f}\Big(\Delta\sigma^{\gamma
q}(x)\otimes\Delta q_i(x,\u^2)+\gg_{\rm hard}(x,Q^2,\u^2)\otimes\Delta
G(x,\u^2)\Big).
\en
That is, the hard piece of $\gg(x)$ contributes to $g_1^p(x)$, while the
soft part is factorized into the nonperturbative quark spin densities
$\Delta q_i(x)$. Since $\Delta\sigma^{\gamma p}(x)$ is a physical quantity,
a different factorization scheme amounts to a different way of shifting the
contributions between $\gg_{\rm hard}(x)$ and $\Delta q(x)$. An obvious
partition of $\gg(x)$ is that the region where $k_\perp^2\gsim\u^2$
contributes to the hard cross section, whereas the soft part receives
contributions from $k_\perp^2\lsim\u^2$ and hence can be interpreted
as the quark and antiquark spin densities in a gluon, i.e.,
$\gg_{\rm soft}(x,\u^2)=\Delta q^G(x,\u^2)$.
Physically, the quark and antiquark jets produced in deep inelastic
scattering with $k_\perp^2\lsim \u^2$ are not hard enough to satisfy the
jet criterion and thus should be considered as a part of one-jet cross
section [12]. The choice of the ``ultraviolet" cutoff for soft contributions
specifies the factorization
convention. In the present paper we only focus on two extremes: the
chiral-invariant scheme in which the ultraviolet regulator respects chiral
symmetry, and the gauge-invariant scheme in which gauge symmetry is respected
but chiral symmetry is broken by the cutoff.

    For a massless quark, one will expect that, based on helicity conservation
or chiral symmetry, the quark spin $\Delta q=\int^1_0\Delta q(x)dx$
is $Q^2$ independent and that there is no sea polarization perturbatively
induced by hard gluons. One way of calculating $\gg_{\rm soft}(x)$ is to
make a direct cutoff on the $k_\perp$ integration so that its integral
expression is exactly the same as $\gg(x,Q^2)$ except that $k_\perp^2$
is integrated over from 0 to $Q^2(1-x)/4x$ for the latter, but
from 0 to $\u^2$ for the former. For $\u^2>>m^2,~-p^2$,
the results are [4,15] (for a complete expression of $\gg_{\rm soft}(x)$ to
NLO using mass or momentum regulator, see [16])
\be
\gg_{\rm soft}(x,\u^2)_{\rm CI}=\Delta q^G_{\rm CI}(x,\u^2)=\cases{
{\alpha_s\over
2\pi}\left[(2x-1)\ln{\u^2\over m^2-p^2x(1-x)}+(1-x)\,{2m^2-p^2x(1-2x)\over
m^2-p^2x(1-x)}\right];   \cr
{\alpha_s\over 2\pi}\left[(2x-1)\ln(\u^2/\ums^2)+2(1-x)\right],   \cr}
\en
for various soft cutoffs, and the subscript CI indicates that we are working
in a chiral-invariant factorization scheme. Note that, as stressed in [9],
the soft cross sections or quark spin densities in a helicity $+$ gluon given
by (9) do not make sense in QCD as they are derived using perturbation theory
in a region where it does not apply. Nevertheless, it is instructive to
see that $\Delta q^G_{\rm CI}=\int^1_0\Delta q^G_{\rm CI}(x)dx$ vanishes
when $m^2=0$ or $-p^2>>m^2$, as expected. Hence, a sea polarization for
massless quarks, if any,
is produced nonperturbatively. Now it does make sense in
QCD to subtract $\gg_{\rm soft}$ from $\gg$ [see Eqs.(4-6)] to obtain a
reliable perturbative QCD result for $\gg_{\rm hard}$:
\be
\gg_{\rm hard}(x,Q^2,\u^2)_{\rm CI}=\,{\alpha_s\over 2\pi}(2x-1)\left
(\ln{Q^2\over\u^2}+\ln{1-x\over x}-1\right),
\en
which is independent of the infrared regulators as long as
$\u^2>>\ums^2,m^2,-p^2$. It is also clear that the soft $2(1-x)$ term in
(5) and (6) drops out in $\gg_{\rm hard}(x)$. Therefore,
\be
\gg_{\rm hard}(Q^2,\u^2)_{\rm CI}=\int^1_0dx\gg_{\rm hard}(x,Q^2,\u^2)_{\rm
CI}=-{\alpha_s\over 2\pi}.
\en
Since gauge invariance and helicity conservation in the quark-gluon vertex
are not broken in the chiral-invariant factorization scheme, it is evident
that $\Delta q_{\rm CI}$ does not evolve, consistent with the naive
intuition, and it corresponds to a nucleon matrix element of a gauge-invariant
but nonlocal operator (see e.g., Eq.(4.7) of [4]).

    Nevertheless, we do have freedom to redefine $\gg_{\rm hard}(x)$ and
$\Delta q(x)$ in accord with Eq.(8). The fact that there is no gluonic operator
at the twist-2, spin-1 level in the approach of OPE [17] indicates that there
must exist a factorization scheme in which hard gluons make no contribution
to the first moment of $g_1^p(x)$ and that $\Delta q$ can be expressed as
a nucleon matrix element of a local gauge-invariant operator. In this scheme,
gluons can induce a sea polarization even for massless quarks. This can be
implemented as follows. For $k_\perp^2\lsim\u^2$, the box diagram for
photon-gluon scattering is reduced under the collinear approximation for the
quark-antiquark pair created by the gluon to a triangle diagram
with the light-cone cut vertex $\gamma^+\gamma^5$ combined
with a trivial photon-quark scattering [4]. As a result, $\Delta q^G(x)$  can
be also obtained by calculating the triangle diagram with the
constraint $k_\perp^2\lsim \u^2$. In order to have a non-vanishing $\Delta q
^G$ even for massless quarks, evidently we need to integrate over
$k_\perp^2$ from 0 to $\infty$ to achieve the axial anomaly and hence
chiral-symmetry breaking [12], and then
identify the ultraviolet cutoff with $\u$. We see that the desirable
ultraviolet regulator must be gauge-invariant but chiral-variant owing
to the presence of the QCD anomaly in the triangle diagram. Obviously,
the dimensional and Pauli-Villars regularizations, which respect the axial
anomaly, are suitable for our purpose. It is found [18]
\footnote{For a complete expression of $\Delta q^G_{\rm GI}(x)$, see [16].
Note that the r.h.s. of Eq.(9) in [16] is too large by a factor of 2.}
\be
\Delta q^G_{\rm GI}(x,\u^2)-\Delta q^G_{\rm CI}(x,\u^2)=-{\alpha_s
\over \pi}(1-x)
\en
for $\u^2>>\ums^2,m^2,-p^2$, where the subscript GI designates the gauge
invariant scheme. It should be accentuated that the last term in (12)
originating from the axial anomaly [12] comes from the region
$k_\perp^2\sim\u^2$.
As noted in passing, the quark spin distribution in a gluon
cannot be reliably calculated by perturbative QCD; however, the difference
between $\Delta q^G_{\rm GI}(x)$ and $\Delta q^G_{\rm CI}(x)$ is trustworthy
in QCD. It is interesting to see from Eq.(12) that
\be
\Delta q^G_{\rm GI}(\u^2)=-{\alpha_s(\u^2)\over 2\pi}~~~~~~{\rm
for~massless~quarks},
\en
that is, the sea-quark polarization perturbatively generated by helicity $+$
hard gluons via the anomaly is {\it negative} !
It follows that the hard cross section has the form
\be
\gg_{\rm hard}(x,Q^2)_{\rm GI} &=& \gg_{\rm hard}(x,Q^2)_{\rm CI}+{\alpha_s
\over \pi}(1-x)   \non \\
&=& {\alpha_s\over 2\pi}\left[(2x-1)\left(\ln{Q^2
\over \u^2}+\ln{1-x\over x}-1\right)+2(1-x)\right],
\en
and its first moment vanishes:
\be
\int^1_0dx\gg_{\rm hard}(x,Q^2)_{\rm GI}=0
\en
in the gauge-invariant factorization scheme. Bodwin and Qiu [4] have shown
generally that the gluonic contribution to $\Gamma_1^p$ vanishes so long
as the ultraviolet regulator for the spin-dependent quark distributions
respects gauge invariance and the analytic structure of the unregulated
distributions. From Eqs.(9) and (12) we also
see that, contrary to what has been often claimed in the literature [7-9],
though the $2(1-x)$ term in (5) and (6) drops out in $\gg_{\rm hard}(x)_{
\rm CI}$ because it arises from the
soft region $k_\perp^2\sim m^2,\ums^2$, it emerges again in the gauge
invariant scheme due to the axial anomaly and this time reappears in the hard
region $k^2_\perp\sim \u^2$.
Therefore, the hard gluonic coefficient function is genuinely {\it hard} !

  We wish to stress that the quark spin density $\Delta q^G(x)$ measures
the polarized sea quark distribution in a helicity $+$ gluon rather than in a
polarized proton.
Consequently, $\Delta q^G(x)$ must convolute with $\Delta G(x)$ in order to be
identified as the sea quark spin distribution in a proton:
\be
\Delta q^{\rm GI}_s(x,\u^2)-\Delta q^{\rm CI}_s(x,\u^2)=-{\alpha_s\over
\pi}(1-x)\otimes\Delta G(x,\u^2).
\en
This relation can be also derived from Eqs.(8) and (14) by noting that
$\Delta\sigma^{\gamma q}(x)=e_q^2\delta(x-Q^2/2p\cdot q)$ [19]. Now it
becomes clear from Eqs.(2), (14) and (16) that {\it despite of the
factorization-scheme dependence of $\Delta q(x)$ and $\gg_{\rm hard}(x)$,
the physical quantity $g_1^p(x)$ remains to be scheme independent up to NLO},
as it should be. The choice of factorization is thus a matter of convention.
Since the valence quark spin distribution $\Delta q_v(x)=\Delta q(x)-
\Delta q_s(x)$ is factorization independent, it follows that [19]
\be
\Delta q_{\rm GI}(x,\u^2)-\Delta q_{\rm CI}(x,\u^2)=-{\alpha_s\over
\pi}(1-x)\otimes\Delta G(x,\u^2),
\en
which leads to
\be
\Delta q_{\rm GI}(Q^2)-\Delta q_{\rm CI}(Q^2)=-{\alpha_s(Q^2)\over 2\pi}\Delta
G(Q^2),
\en
where we have set $\u^2=Q^2$.
Eqs.(14) and (17) provide the necessary relations between the gauge-invariant
and chiral-invariant factorization schemes. For a given $\Delta G(x,Q^2)$,
the quark spin densities in these two different schemes are related to each
other via (17). It should be remarked that in spite of a vanishing gluonic
contribution to $\Gamma_1^p$ in the gauge-invariant scheme,
\footnote{A sea-qaurk interpretation of $\Gamma_1^p$ with $\Delta s=-0.10
\pm 0.03$ at $Q^2=10\,{\rm GeV}^2$ [20] has been
criticized on the ground that a bound $|\Delta s|\leq
0.052^{+0.023}_{-0.052}$ [21] can be derived based on the information of
the behavior of $s(x)$ measured in deep inelastic neutrino
experiments and on the positivity constraint. First of all, this argument is
quite controversial [22]. Second, one can always find a polarized strange
quark distribution with $\Delta s\sim -0.10$ which satisfies positivity
and experimental
constraints [3]. Third, a sea polarization of order $-0.11$
is also found by lattice calculations [23,24].}
it never means that $\Delta G$ vanishes in a polarized proton.
Quite opposite to the naive expectation, if there is no sea polarization
in the chiral-invariant scheme, then the size of the gluon spin component in a
proton must numerically obey the relation
$\Delta G(Q^2)=-(2\pi/ \alpha_s(Q^2))\Delta q_s^{\rm GI}(Q^2)$
in order to perturbatively generate a negative sea-quark polarization
$\Delta q_s^{\rm CI}(Q^2)$ via the QCD anomaly. In other words, even
gluons do not contribute to $\Gamma_1^p$, the gluon spin can be as large as
2.5 for $\Delta q_s^{\rm GI}=-0.10\,$ at $Q^2=10\,{\rm GeV}^2$ provided
that $\Delta q_s^{\rm CI}=0$. Recall that the gluon polarization induced
from quark's bremsstrahlung is positive (see the first moment of $\Delta P_{Gq}
(x)$ in Eq.(25) below).

   Phenomenologically, one has to specify the factorization scale $\u$ in
order to extract the quark and gluon spin distributions from polarized DIS
data as the hard photon-gluon cross section is dependent of $\u$. In practice,
$\u^2$ can be fixed to be $\la Q^2\ra$, the average $Q^2$ of the $g_1^p(x)$
data set. Of course, one should take into consideration the logarithmic term
$\ln(Q^2/\u^2)$ in $\gg_{\rm hard}(x)$  to fully account for
the measured $x$ dependence of $Q^2$ at each $x$ bin.

   One may ask how to accommodate the aforementioned two different
factorization schemes
in the approach of OPE ? An examination of this issue also provides a clear
picture on the differences between $\Delta q_{\rm GI}$ and $\Delta q_{\rm
CI}$. Since the flavor-singlet axial-vector current $J_\mu^5$ has an
anomalous dimension first appearing at the two-loop level [25], the
quark spin
\be
\Delta q_{\rm GI}=\,\la p|\bar{q}\gamma_\mu\gamma_5q|p\ra s^\mu,
\en
with $s^\mu$ a spin 4-vector, is gauge-invariant but $Q^2$ dependent. The
evaluation of the nucleon matrix element of $J_\mu^5$ involves
connected and disconnected insertions (see e.g., [23]). The connected
and disconnected insertions are related to valence quark and vacuum (i.e.,
sea quark) polarizations, respectively, and are separately gauge invariant.
Thus we can make the identification:
\be
\la p|J_5^\mu |p\ra=\la p|J_5^\mu|p\ra_{\rm con}+\la p|J_5^\mu|p\ra_{\rm dis}
=\sum_q(\Delta q^{\rm GI}_v+\Delta q_s^{\rm GI})s^\mu.
\en
Interestingly, lattice QCD calculations of $\Delta q^{\rm GI}_v$ and $\Delta
q^{\rm GI}_s$ became available very recently [23,24]. It is found that
$\Delta u_s=\Delta d_s=\Delta s=-0.12\pm 0.01$ from the disconnected
insertion [23]. This empirical SU(3)-flavor symmetry within errors
implies that the sea-quark polarization in the gauge-invariant scheme
is indeed predominately generated by the axial anomaly. Recall that
sea contributions in the unpolarized case are far from being SU(3) symmetric:
$\bar{d}>\bar{u}>\bar{s}$. In order to connect to the chiral-invariant
scheme, one can write
\be
J^\mu_5=\,J^\mu_5-K^\mu+K^\mu\equiv\,\tilde{J}^\mu_5+K^\mu,
\en
with $K^\mu=(\alpha_sn_f/2\pi)\epsilon^{\mu\nu\rho\sigma}A^a_\nu(\partial
_\rho A^a_\sigma-{1\over 3}gf_{abc}A_\rho^bA_\sigma^c)$ and $\epsilon_{0123}
=1$. Though neither $\tilde{J}^\mu_5$ nor $K^\mu$ is gauge invariant,
their matrix elements can be identified with quark and gluon spin components
in the light-front gauge $A^+=0$ [12] (it is not necessary
to specify the coordinate)
\be
s^+\Delta q_{\rm CI}=\la p|\tilde{J}^+_5|p\ra_{A^+=0},~~~~s^+\Delta G=
\la p|(\vec{E}\times\vec{A})^+|p\ra_{A^+=0}=
-{2\pi\over \alpha_sn_f}\la p|K^+|p\ra_{A^+=0}.
\en
The quark spin $\Delta q_{\rm CI}$ does not evolve as the current
$\tilde{J}^\mu_5$ is conserved in the chiral limit. Of course, both
$\Delta G(x)$ and $\Delta q_{\rm CI}(x)$ (not just their first moments!)
also can be recast as matrix elements of a
gauge-invariant but nonlocal operator [4,15], as noted in passing. Applying
(22) to the axial-current matrix element leads to
\be
\la p|J_5^\mu |p\ra &=& \la p|J_5^\mu|p\ra_{\rm con}+\la p|\tilde{J}_5
^\mu|p\ra_{\rm dis}+\la p|K^\mu|p\ra_{\rm dis}   \non \\
& \buildrel A^+=0 \over\longrightarrow & \sum_q(\Delta q^{\rm CI}_v+
\Delta q_s^{\rm CI}-{\alpha_s\over 2\pi}\Delta G)s^+,
\en
where use of $\Delta q^{\rm GI}_v=\Delta q^{\rm CI}_v$ has been made. It is
clear that (20) and (23) are equivalent owing to the relation (18).

\vskip 0.3cm
{\bf 3.}~~The $Q^2$ dependence of parton spin densities are governed
by the Altarelli-Parisi equations:
\be
&& {d\over dt}\Delta q_{\rm NS}(x,t)=\,{\alpha_s(t)\over 2\pi}\Delta P_{qq}^
{\rm NS}(x)\otimes\Delta q_{\rm NS}(x,t),   \non \\
&& {d\over dt}\left(\matrix{\Delta q_{\rm S}(x,t)   \cr   \Delta G(x,t) \cr}
\right)=\,{\alpha_s(t)\over 2\pi}\left(\matrix{\Delta P_{qq}^{\rm S}(x) &
2n_f\Delta P_{qG}(x)  \cr  \Delta P_{Gq}(x) & \Delta P_{GG}(x) \cr}
\right)\otimes\left(\matrix
{\Delta q_{\rm S}(x,t)   \cr   \Delta G(x,t) \cr} \right),
\en
where $\Delta q_{\rm NS}(x)=\Delta q_i(x)-\Delta q_j(x),~\Delta q_{\rm S}(x)=
\sum_i\Delta q_i(x)$ and $t=\ln(Q^2/Q^2_0)$. The complete polarized
splitting functions up to NLO, $\Delta P(x)=\Delta P^{(0)}(x)+{\alpha_s\over 2
\pi}\Delta P^{(1)}(x)$, have been calculated in the $\overline{\rm MS}$ scheme
recently [5]. Hence, the NLO evolution of spin parton distributions
in the gauge-invariant factorization scheme is completely determined.
Explicitly, the AP equation for the first moment of flavor-singlet parton spin
densities reads [5]
\be
{d\over dt}\left(\matrix{\Delta\Sigma_{\rm GI}(t) \cr \Delta G(t)  \cr}\right)
=\,{\alpha_s(t)\over 2\pi}\left(\matrix{ {\alpha_s\over 2\pi}(-2n_f) & 0  \cr
2+{\alpha_s
\over 2\pi}(25-{2\over 9}n_f)  &  {\beta_0\over 2}+{\alpha_s\over 2\pi}\,{
\beta_1\over 4}   \cr}\right)\left(\matrix{ \Delta\Sigma_{\rm GI}(t)  \cr
\Delta G(t)\cr}\right),
\en
where $\Delta\Sigma_{\rm GI}(t)=\int^1_0dx\Delta q_{\rm S}(x,t),~\beta_0=11-{
2n_f\over 3}$ and $\beta_1=102-{38\over 3}n_f$. It is clear that $\Delta
\Sigma_{\rm GI}$ to NLO is $Q^2$ dependent.

One may choose to work in the chiral-invariant factorization scheme, so that
$\Delta\Sigma_{\rm CI}$ does not evolve with $Q^2$. This requires that
\be
\gg_{\rm hard}(Q^2)_{\rm CI}=-{\alpha_s\over 2\pi},~~~~\gamma_{qq}^{(1)S,1}
\equiv \int^1_0\Delta P_{qq}^{(1)S}(x)dx=0.
\en
Using the results obtained in the $\overline{\rm MS}$ scheme, one may
introduce a modification on NLO anomalous dimensions and hard coefficient
functions to transfer from the GI scheme to the CI prescription [26]. However,
this
transformation cannot be unique since it is only subject to the constraints
(26). Indeed, three different scheme changes have been constructed in [7].
As a consequence, the NLO evolution of polarized parton distributions in the
CI scheme obtained in this manner [7] is ambiguious and not quite trustworthy
as it depends on the scheme of transformation.

   We can avoid the aforementioned ambiguities and complications by
working in the context of the gauge-invariant scheme where NLO polarized
splitting functions are known. Once the evolution of the spin parton
distributions $\Delta q_{\rm GI}(x,Q^2)$ and $\Delta G(x,Q^2)$
is determined from the AP equations, $\Delta q_{\rm CI}(x,Q^2)$ in
the chiral-invariant prescription is simply related to $\Delta q_{\rm GI}(x,
Q^2)$ and $\Delta G(x,Q^2)$ by Eq.(17). One can check from Eq.(25) that
$d(\Delta\Sigma_{\rm CI}(t))/dt=0$ to NLO, as it should be, where $\Delta
\Sigma_{\rm CI}(t)=\Delta\Sigma_{\rm GI}(t)+(n_f\alpha_s/2\pi)\Delta G(t)$.
In the absence of direct calculations of NLO polarized splitting functions
in the chiral-invariant scheme, we believe that this is the right approach
for studying the $Q^2$ evolution of parton spin distributions.

\vskip 0.3cm
{\bf 4.}~~To conclude, contrary to what has been often claimed in the
literature, we
have clarified that the $2(1-x)$ term in $\gg_{\rm hard}(x)$ in the
gauge-invariant factorization scheme arises from the region
$k_\perp^2\sim\u^2$ and hence is a genuinely hard contribution.
We have shown explicitly that the physical quantity
$g_1^p(x)$ is independent of the choice of factorization convention. The
sea-quark interpretation of $\Gamma_1^p$ in the gauge-invariant scheme or
the anomalous gluon interpretation in the chiral-invariant scheme is
purely a matter of factorization convention chosen in defining $\Delta q(x)$
and $\gg_{\rm hard}(x)$. We have emphasized that even
though hard gluons do not contribute to $\Gamma_1^p$ in the gauge-invariant
scheme, the gluon spin component in a proton should be large enough to
perturbatively generate a negative sea polarization via the axial anomaly,
recalling that $\Delta G(x,Q^2)$ is factorization independent.

    As far as $g_1^p(x)$ is concerned, both GI and CI factorization schemes
are on the same footing. Thus it does not make sense to keep
disputing which factorization prescription is correct or superior.
Of course, once a set of $\Delta q_{\rm GI}(x),~\Delta G(x),~\gg
_{\rm hard}(x)_{\rm GI}$ or of $\Delta q_{\rm CI}(x),~\Delta G(x),~\gg_{\rm
hard}(x)_{\rm CI}$ is chosen, one has to stick to the same scheme in all
processes.

  In practice, it appears that the use of $\Delta q_{\rm GI}(x)$ is more
convenient than $\Delta q_{\rm CI}(x)$. First of all, $\Delta q_{\rm GI}$
corresponds to a nucleon matrix element of a local and gauge-invariant
operator, and its
calculation in lattice QCD became available recently. For $\Delta q_{\rm CI}$,
one has to compute the matrix element of $\tilde{J}^\mu_5$ in the light-front
gauge, which will involve much more lattice configurations. Second, NLO
polarized splitting functions have been determined very recently in
the gauge-invariant scheme, and it is straightforward to study the evolution
of $\Delta q_{\rm GI}(x,Q^2)$ through AP evolution equations.  We have
argued that a NLO analysis of polarized DIS data should be first carried
out in the gauge-invariant factorization scheme and then related to
the chiral-invariant prescription, if desired, via Eq.(17).

\vskip 2.0cm
\centerline{\bf ACKNOWLEDGMENTS}
\vskip 0.3 cm
    I wish to thank H.L. Yu for a careful reading of the manuscript.
    This work was supported in part by the National Science Council of ROC
under Contract No. NSC85-2112-M-001-010.

\vskip 1.5 cm
\centerline{\bf REFERENCES}
\vskip 0.3 cm
\begin{enumerate}

\item EMC Collaboration, J. Ashman {\it et al.,} \np {\bf B238}, 1 (1990); \pl
{\bf B206}, 364 (1988).

\item SMC Collaboration, D. Adams {\it et al.,} \pl {\bf B329}, 399 (1994);
{\bf B339}, 332(E) (1994).

\item H.Y. Cheng, H.S. Liu, and C.Y. Wu, IP-ASTP-17-95 [hep-ph/9509222]
(revised version), to appear in Phys. Rev. D.

\item G.T. Bodwin and J. Qiu, \pr {\bf D41}, 2755 (1990), and
in {\it Proc. Polarized Collider Workshop}, University Park, PA, 1990, eds.
J. Collins {\it et al.} (AIP, New York, 1991), p.285.

\item R. Mertig and W.L. van Neerven, INLO-PUB-6/95 [hep-ph/9506451];
W. Vogelsang, RAL-TR-95-071 [hep-ph/9512218].

\item M. Gl\"uck, E. Reya, M. Stratmann, and W. Vogelsang, DO-TH 95/13
[hep-ph/9508347]; W. Vogelsang, RAL-TR-95-059 [hep-ph/9510429].

\item R.D. Ball, S. Forte, and G. Ridolfi, CERN-TH/95-266 [hep-ph/9510449];
R.D. Ball, Edinburgh 95/558 [hep-ph/9511330]; S. Forte,
CERN-TH/95-305 [hep-ph/9511345].

\item W. Vogelsang, \zp {\bf C50}, 275 (1991).

\item L. Mankiewicz, \pr {\bf D43}, 64 (1991).

\item A.V. Manohar, in {\it Proc. Polarized Collider Workshop}, University
Park, PA, 1990, eds. J. Collins {\it et al.} (AIP, New York, 1991), p.90.

\item R.D. Carlitz and A.V. Manohar, in {\it Proc. Polarized Collider
Workshop}, University Park, PA, 1990, eds. J. Collins {\it et al.} (AIP, New
York, 1991), p.377.

\item R.D. Carlitz, J.C. Collins, and A.H. Mueller, \pl {\bf B214}, 229
(1988).

\item G. Altarelli and G.G. Ross, \pl {\bf B212}, 391 (1988).

\item P. Ratcliffe, \np {\bf B223}, 45 (1983).

\item A.V. Manohar, \prl {\bf 66}, 289 (1991).

\item F.M. Steffens and A.W. Thomas, ADP-95-38/T192 [hep-ph/9510460].

\item R.L. Jaffe and A.V. Manohar, \np {\bf B337}, 509 (1990).

\item S.D. Bass, \zp {\bf C55}, 653 (1992);
S.D. Bass and A.W. Thomas, {\sl J. Phys.} {\bf G19}, 925 (1993).

\item H.Y. Cheng and C.F. Wai, \pr {\bf D46}, 125 (1992).

\item J. Ellis and M. Karliner, \pl {\bf B341}, 397 (1995).

\item G. Preparata and J. Soffer, \prl {\bf 61}, 1167 (1988);
{\bf 62}, 1213(E) (1989).

\item J. Soffer, CPT-92-P-2809 (1992).

\item S.J. Dong, J.-F. Laga\"e, and K.F. Liu, \prl {\bf 75}, 2096 (1995).

\item M. Fukugita, Y. Kuramashi, M. Okawa, and A. Ukawa, \prl {\bf 75}, 2092
(1995).

\item J. Kodaira {\it et al.,} \np {\bf B159}, 99 (1979);
\pr {\bf D20}, 627 (1979).

\item E.B. Zijlstra and W.L. van Neerven, \np {\bf B417}, 61 (1994).

\end{enumerate}

\end{document}